\documentclass[pre,superscriptaddress,twocolumn]{revtex4}

\usepackage{graphicx}

\begin{document}

\title{Effects of relaxation processes during deposition of anisotropic
grains  on a flat substrate}

\author{Kamil Trojan} 
\affiliation{SUPRATECS Centre, Institute of Physics, B5,
University of Li$\grave e$ge,\\ B-4000 Li$\grave e$ge, Euroland}
\affiliation{Institute of Theoretical Physics, University of
Wroc\l{}aw,\\  pl. M. Borna 9, 50-204 Wroc\l{}aw, Poland} 
\author{Marcel Ausloos} 
\affiliation{SUPRATECS Centre, Institute of Physics, B5,
University of Li$\grave e$ge,\\ B-4000 Li$\grave e$ge, Euroland}

\begin{abstract}
The ballistic deposition on a one dimensional substrate of grains 
with one degree of freedom, called spin,  is studied with respect to 
relaxation processes during deposition. The ''spin''  represents the 
grain anisotropy, e.g. its longest axis with respect to the vertical. 
The grains interact through some contact energy ($J$) and are allowed 
to flip with a probability $q$ during deposition and relaxation. 
Different relaxation processes are investigated. The pile structure 
is investigated, i.e. the density and ''magnetisation'', as a 
function of $q$ and $J$. A percolation transition is found across 
which the cluster size changes from exponential-like to a power 
law-like dependence. The differences between ''ferromagnetic'' and 
''anti-ferromagnetic''-like contact energies are emphasized as a 
function of $q$.
\end{abstract}

\maketitle

\section{ INTRODUCTION}

The sandpile problem \cite{bakbook} is one of the most recently often
tackled problems in the self-organisation of complex systems. There are
several questions of interest, not only concerning the density,
compaction, stability, intrinsic dynamics, etc. but also concerning the
detailed non equilibrium physics inherent to such systems
\cite{reposeangle2}. Experimental observations pertain to the force
structure \cite{dutchguy1} in the pile and to the clustering or 
percolation properties of grain piles \cite{cuba}.
It is known that the system is hyperstatic
\cite{dutchguy1}. In fact the systems are not at all  in equilibrium in a
thermodynamic sense, resulting in hysteresis and ageing  processes. A
description of the construction of the pile is thus  a relevant input
before further studies.

One important physical constraint to be  considered in describing
granular piles is that  the materials are not made of symmetrical
(spherical or cubic) entities\cite{mehta}. Whence grains  can be imagined to be 
identical entities but having one degree of freedom, call it a {\em spin},
for usual statistical mechanics considerations.   Such a ''spin'' allows
us for referring to a direction or a rotation process. The spin
indicates, e.g. the orientation of the longest grain axis with respect
to the vertical. 

We have studied such a model \cite{our,ijmpc}, within a (magnetic)
ballistic deposition process; it is similar to the Tetris model
\cite{tetris}, but more simplified. Basically the  (2-dimensional)
grains  are anisotropic and are deposited vertically as in a gravity
controlled rain. For simulation studies we have imagined a two
dimensional vertical silo like  a triangular lattice with vertical
edges, but with periodic boundary conditions. The grains can be
submitted to an external field during deposition such that their
orientation can change (or not) at each step, with a certain
probability. The grain energy is calculated at each falling down step
according to a sort of Ising energy, the exchange integral being a map
of the contact energy between grains. The latter can have various
origins: surface roughness, local charges, mechanical, chemical, dipolar
or magnetic effects, ...  . It is known that (long range) magnetic
interactions lead to clustering and aggregation \cite{MDipolDLA}. For
simulation ease, we can limit the interactions to a short range.

We have previously examined  clustering effects as in \cite{our,ijmpc},
through the cluster fractal dimension, the pile overall density, the
distribution of grain orientations, ... and have found through
percolation arguments \cite{Staufferbook,pekal} the existence of different
cluster growth ''mechanisms''.

However, except for a Metropolis like dynamical condition
\cite{Metropolis} during the Monte Carlo simulation, we have not fully
let the system relax during the deposition.  This constraint is hereby
removed. We have performed simulations under two pile relaxation
conditions to be described in Sect. \ref{sec:exper}. One is a total gravitationally
controlled relaxation, the other is a blocked gravitationally controlled relaxation,
relaxing the ballistic constraint, and allowing for the grain to move to
a neighboring column. It should be expected that such a relaxation has a
coarsening  effect

The results are found in Sect. \ref{sec:numres} while  Sect. \ref{sec:conclus} 
contains some brief
conclusion

\section{ EXPERIMENTAL PROCEDURE}\label{sec:exper}

A few changes to the experimental procedure presented in \cite{our} are
hereby introduced. We propose an algorithm for creating a pile 
with a fixed  probability $q
\in [0;1]$ for spin flip. The algorithm for arbitrary $q$ (so called
$q$-MBD model, in contrast to the $\frac{1}{2}$-MBD model \cite{our})
goes as follows:

\begin{enumerate}

\item first, we choose a 1-dimensional crenellated horizontal substrate of spins
with a predetermined (hereby random) configuration; periodic boundary
conditions are used; on each site a vertical column of empty cells is placed. The lattice
symmetry is thus hexagonal.

\item{\label{en:step2}} a falling (up or down) spin is dropped along one of
the lattice columns from a height $r_{max}+5a$, where $r_{max}$ is the
largest distance between an occupied cluster site and the substrate,
i.e. the height of the highest column growing from the substrate on  the
lattice

\item  more importantly at each step down the spin can flip i.e., change
its ''$sign$'', with probability $q$;  more specifically the probability
to choose the ''up'' direction is $q$;

\item the spin goes down flipping until it reaches a site perimeter of
the cluster at which time  the local gain in the Ising energy (in
absence of an external field)

\begin{equation}
\label{eq:isingham} 
\beta E = -\beta J \sum_{<i,j>} \sigma_i \sigma_j, 
\end{equation}

thus for possible cluster growth is calculated. The exchange integral
$J$  describing spin-spin interaction can be thought in granular matter
to be like some contact energy due to the surface roughness or the
elastic, electric, magnetic, chemical ... character of grains; it can be
also a measure of some misorientation entropy. If the energy gain is
negative the  spin sticks to the cluster immediately (sticking
probability =1.0) and one goes  back to step (\ref{en:step2}). In the
opposite case the  spin sticks to the cluster with a rate $\exp (-\Delta
\beta E)$ where $\Delta \beta E$ is the local gain in the Ising energy.
If the spin does not stick to the cluster it continues going down under 
the same stick and fall rules. At some step the spin sticks to the cluster. 
If the site just below the falling spin is occupied the spin is allowed to 
relax by moving down in one of the nearest columns.
 
\item At this point we introduce and study three kinds of relaxation
processes: 

\begin{enumerate} 

\item[I.] {\it No relaxation} -- the no-relaxation process is the 
pure $q$-MBD, \cite{ijmpc}, i.e. the spin
sticks to the cluster immediately. 

\item[II.] {\it Total gravitationally controlled relaxation} -- in the total 
gravitational relaxation process, when the
spin reaches the cluster, i.e. the site just below the spin is occupied,
the spin can, with equal probability: stay on the site, or roll down all the
way in the left or the right neighbouring column if one of the first
neighboring  sites is unoccupied. The spin continues rolling down 
until it reaches a site from which the rolling is not
possible; one assumes that the
rolling process is due to the gravity without any resistance. 

\item[III.] {\it Blocked gravitationally controlled relaxation} -- after reaching the
cluster the spin can, with equal probability: stay  on the site, roll
down on the left or on  the right from the site as for the total
gravitational relaxation (II), but can also stick to the pile on its way
to the lowest level: During every move towards  the lowest level the
falling spin can be stopped and stick to the cluster depending on the
local gain in the Ising energy (\ref{eq:isingham}). 
\end{enumerate}

If the spin cannot move (due to the relaxation), it immediately stops
and sticks to the cluster.

\item After the spin sticks to the cluster one goes back to step
(\ref{en:step2}). 

\end{enumerate}

After dropping a (large) number of spins the physical quantities of
interest like magnetization, density, histogram of up/down spin cluster
size, histogram of hole cluster size, etc.  are computed.

\section{ NUMERICAL RESULTS}

\label{sec:numres} All results reported below are for a triangular lattice
with vertical edges
of horizontal size $L=100$, i.e. the width of the seed substrate,  and
when the pile made of clusters has reached a $100$ lattice unit height.
The substrate consists of spins with either direction chosen with
probability $q$. Every reported data point hereby corresponds to an
average over $1000$ simulations.

\subsection{ Density} \label{sec:density}

We define the density of a cluster as $$\rho = \frac{\mbox{number of
spins in the cluster}}{\mbox{number of sites on the lattice}},$$ in
which obviously the number of lattice sites  in the denominator $=10
000$.

\begin{figure*} 
\includegraphics[height=16.5cm,angle=-90]{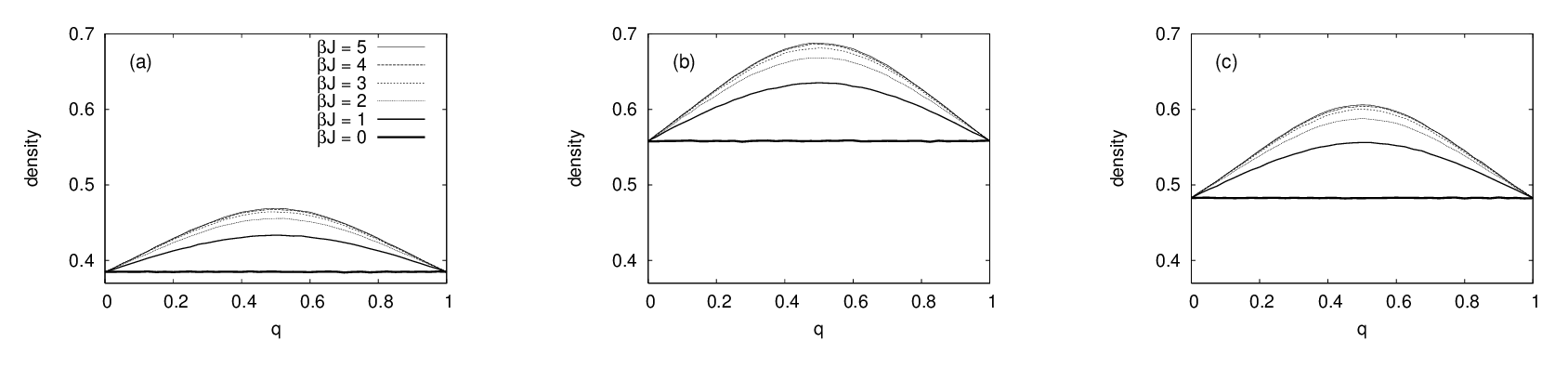} 
\caption{\label{fig:roqf} Projections of the total spin (or grain) density dependence on
$q$ and $\beta J$ where $\beta J>0$: (a) case I, (b) case II, (c) case III.} 
\end{figure*}

\begin{figure*} 
\includegraphics[height=16.5cm,angle=-90]{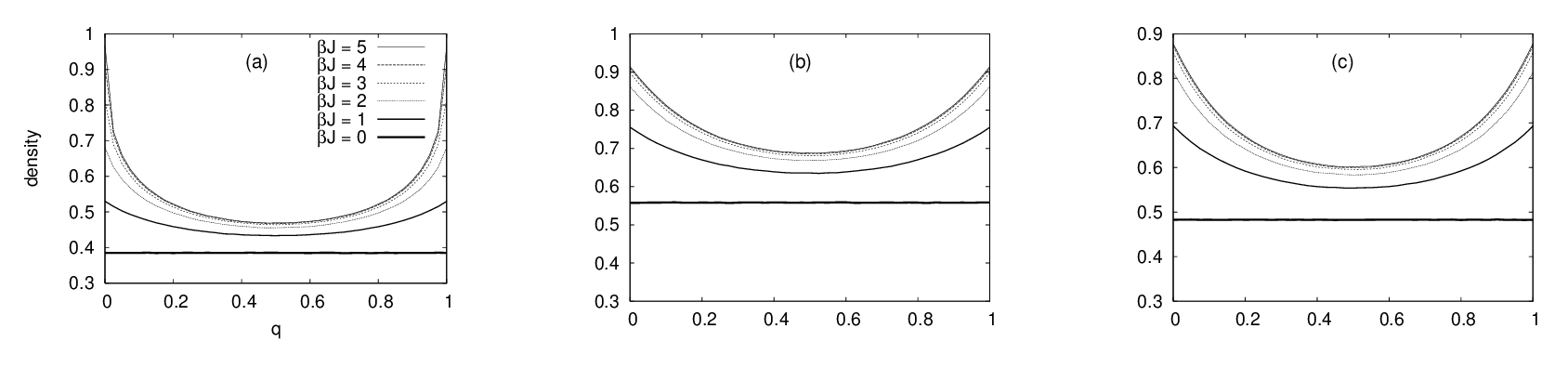} 
\caption{\label{fig:roqaf} Projections of the total spin (or grain) density dependence
on $q$ and $\beta J$  where $\beta J>0$: (a) case I, (b) case II, (c) case III.} 
\end{figure*} 

Figs.\ref{fig:roqf}-\ref{fig:roqaf} illustrate the behaviour of the
density with respect to the $q$ and $\beta J$ parameters. The figures
convince us that the density reached by the pile in the presence of
relaxation is always higher than in absence of relaxation, at fixed $q$
and $\beta J$ values, as expected. The reached density values are not
trivial though (compare with \cite{faraudo}). It can be noticed, as expected also that $\rho^{II}
>\rho^{III} > \rho^{I} $   whatever the interaction sign.

\subsection{ Magnetization}

\begin{figure*} 
\includegraphics[height=16.5cm,angle=-90]{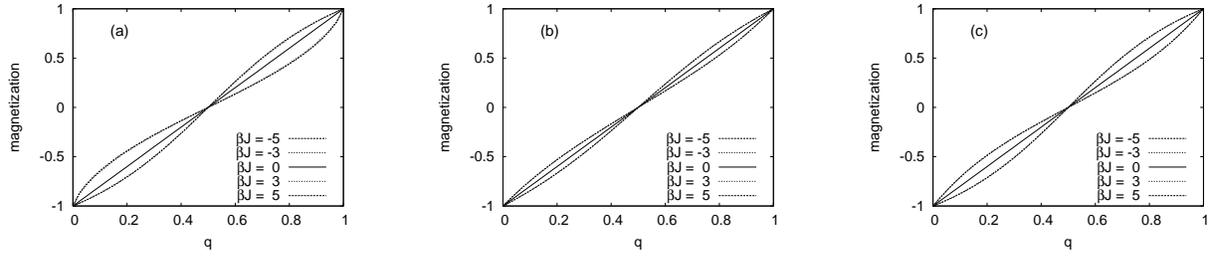}
\caption{\label{fig:mag} The magnetization for 
 (a) the relaxation case I, (b) the relaxation  case II, (c) the relaxation  case III. 
 Three values of $\beta J$ are presented.} 
\end{figure*}

The dependence of the magnetization defined as

\begin{equation} M = \frac{n_+ - n_-}{n_+ + n_-} \end{equation}

is shown in Fig.\ref{fig:mag}(a) as a function of $\beta J$ and $q$,
where $n_+$ and $n_-$ are the number of up and down spins respectively,
i.e. $n_+ = 10000 \rho_+$. This quantity can be considered as  a measure
of the difference in grain orientations in  the packing, when the spin (+)
is understood e.g. as indicating that the grain longest axis orientation is along the vertical.

In \cite{our} a non linear-like dependence on $q$  was found  for
the magnetization though with small departure from linearity due to the changes in the
interaction strength. Notice that the relaxation processes make the
departures from linearity still smaller; the smallest departures exist  in
the case II, total gravitational relaxation (Fig.\ref{fig:mag}b). As it was mentioned in
\cite{our} this has to be expected if one recalls that $q$, in a first
approximation, is the fraction of deposited up-spins.

\subsection{ Percolation} 

In this section we present results concerning
the percolating cluster (from the substrate to the top of the pile) of a
typical entity, i.e. the up-spins. In order to do so every pile built by
the algorithm has been checked in order to determine  whether there is a
percolating cluster of up spins. At fixed $q$ and $\beta J$ the fraction
of piles  consisting of this  given type of percolating cluster was
computed. This value is thought to be representative of the critical
percolation $p_c$.

\begin{figure*}
\includegraphics[height=16.5cm,angle=-90]{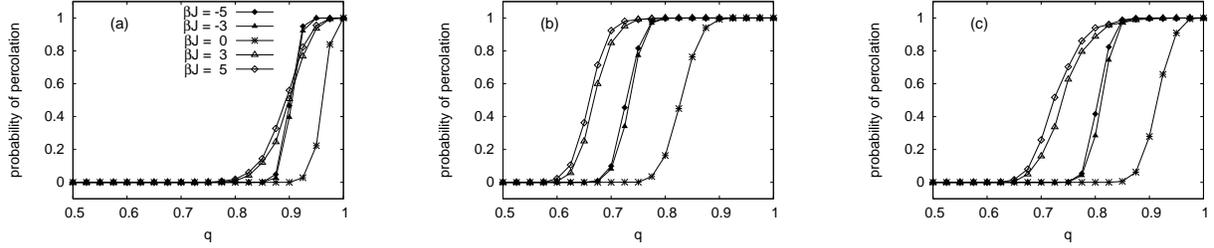}
\caption{\label{fig:percolation} Dependence of the percolation
probability $p_c$ on $q$: 
(a) the relaxation case I, (b) the relaxation  case II, (c) the relaxation  case III. 
Three values of $\beta J$ are presented.} 
\end{figure*}

\begin{figure*}
\includegraphics[height=16.5cm,angle=-90]{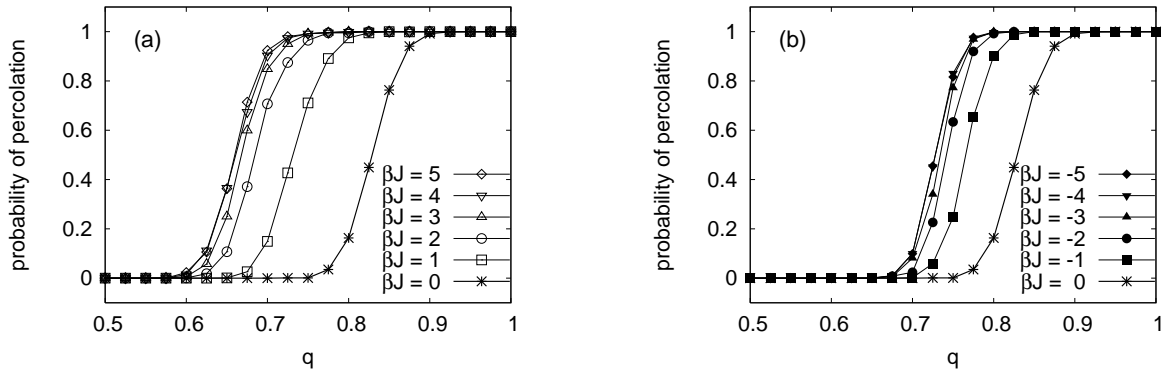}
\caption{\label{fig:percconv} Dependence of the percolation
probability $p_c$ on $q$ for the relaxation  case II (total gravitationally controlled relaxation)
for: (a) $\beta J>0$ and (b) $\beta J<0$.} 
\end{figure*}

Fig.\ref{fig:percolation} shows the behaviour of $p_c$ with respect to
$q$ for several values of the $\beta J$ parameter for the three cases. The sigmoid curve is
likely due to the finite size of the system. Observe that the interval of $q$ for
which $ 0 < p_c < 1$ is wider when relaxation occurs. One can also
observe that the  cases AF or F and  $\beta J=0$ are much more
distinguishable when there is relaxation   (total or blocked gravity).
However some convergence toward some finite $p_c$ value seems to occur
for increasing $|\beta J|$, as seen on Fig.\ref{fig:percconv} for case II, -- the asymptotic 
value depending in each case on $\beta J$ sign and being at a lower $q$ value when $\beta J>0$.

Notice that for $\beta J>0$ in the
relaxation cases percolation occurs for $q > 0.59 = q_0$ (Fig.\ref{fig:percolation}b) while in the
non-relaxation case it exists for $q > 0.8 = q_0$ (Fig.\ref{fig:percolation}a).
Therefore, relaxation has an obvious essential influence on the critical percolation value.
Recall that in a first approximation the $q$ parameter is the percentage of up spins in
the pile, therefore a lower value of $q_0$ means that one does not have
to use a lot of up spins to have a percolation path (this is useful for
example if the up spin material is much more expensive or dangerous).

\subsection{ The size (mass) distribution of clusters}

Many problems in granular matter concern the creation of clusters
according to segregation\cite{segregation}, creation of
avalanches\cite{avalanches} or decompaction\cite{decompactation}. In
this section we present the results of the analysis of the size (mass)
of spin and hole clusters in this  $q$-MBD model with or without
relaxation processes.

\begin{figure*} 
\includegraphics[height=16.5cm, angle=-90]{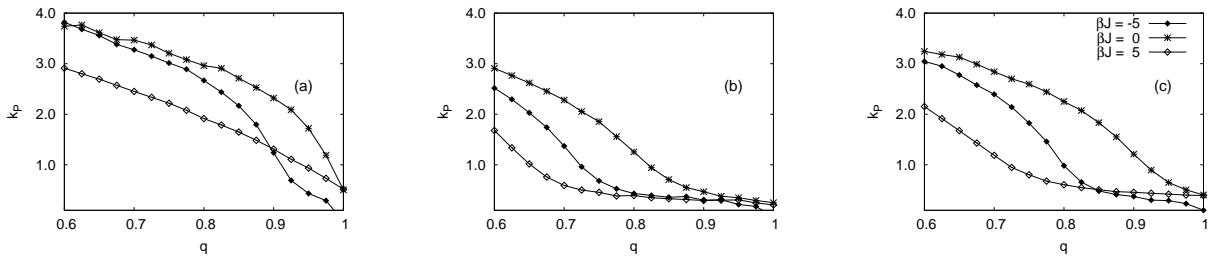} 
\caption{\label{fig:kp}
The exponent $k_p$ form power law regime:
(a) the relaxation case I, (b) the relaxation  case II, (c) the relaxation  case III.} 
\end{figure*} 

In \cite{our} it has been found that the number ($N_s$) of clusters
dependence on the size ($s$) of the clusters demonstrates a crossover
effect, i.e.  from an exponential-like to a power law-like dependence.
The power law-like regime exists for high $q$ values and the exponential
law regime occurs when the $q$ parameter is low. It was conjectured that
the border value is at the critical percolation ($q_c$).

A crossover effect likewise occurs in presence of relaxation. The whole
data analysis is not reproduced here; it was done as for the case I in
\cite{ijmpc}. The differences are  particularly noticeable in the power
law region i.e. for high values of $q$. Fig.\ref{fig:kp} shows the
behaviour of the power law exponent $k_p$, where $N_S \propto e^{-k_ps}$
and $N_s$ is a number of clusters with mass $s$ for the three models.
Notice that the $k_p$ parameter is quasi the same (taking into
consideration error bars) for finite $\beta J$ and $\beta J=0$, when
$q>0.8$, i.e. when $p_c \approx 1$. In that case $k_p$ is almost $q$
independent.

\begin{figure*} 
\includegraphics[height=16.5cm,angle=-90]{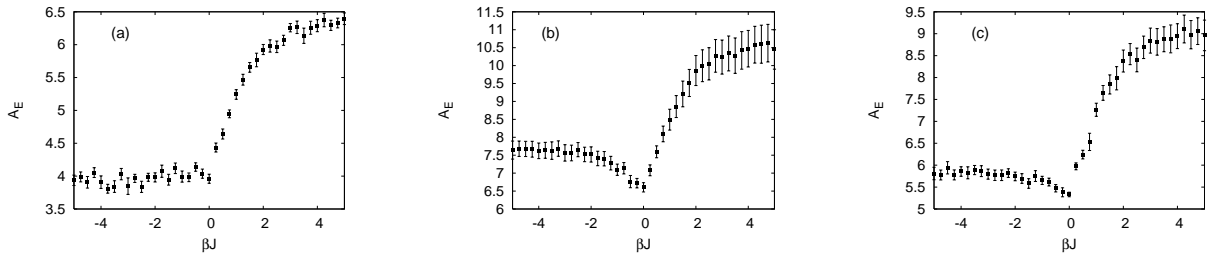} 
\caption{\label{fig:ae}The exponent $A_E$ form exponential law regime:
(a) the relaxation case I, (b) the relaxation  case II, (c) the relaxation  case III.} 
\end{figure*} 

Let us introduce (like in \cite{our}) two parameters to characterize the
exponential law regime: an exponent $k_E$ where $N(s) \sim \exp (-k_E
s)$ and $A_E$ where $k_E = \exp (-A_E q)$. Observe (Fig.\ref{fig:ae}) in
contradistinction with the no relaxation models that $A_E \neq const$ for
$\beta J <0$, but $A_E \cong const$ for $\beta J<-2$. In other words $A_E \cong
const$ for $|\beta J|<2$ in the presence of relaxation, i.e. for small
values of $q$ the cluster growth  does not much depend on the interaction strength 
for higher interaction vaules.

\subsection{ The analysis of spin configurations}

The probability of spin sticking for a given spin neighbourhood
configuration has been counted. The results reveal that  the most
probable configurations  depend on $q$ and $\beta J$.  We have
investigated many configurations in the ($q$, $\beta J$) plane. We
confirm that this approach allows to  estimate as well the percolation
transition, but such results cannot be fully displayed here. 

\begin{table*}
\caption{\label{tab1}The distribution of typical configurations in the 
case (II) for $\beta J=5$; {\em conf. type} corresponds to a configuration
 type (see Fig.\ref{fig:conf}); {\em prob.} corresponds to a probability 
 magnitude for the configuration type. The symmetrical (with respect to 
 vertical axis) configurations are not recognized, i.e.
{\tt 0-000} and {\tt 000-0} are the same configuration types.}
\begin{tabular}{c|l|c|l|c|l|c|l|c|l}
\multicolumn{2}{c|}{q=0} & 
\multicolumn{2}{c|}{q=0.25} & 
\multicolumn{2}{c|}{q=0.5}  & 
\multicolumn{2}{c|}{q=0.75}  & 
\multicolumn{2}{c|}{q=1.0} \\
\hline \hline
conf. type & prob. & conf. type. & prob. & conf. type. & prob. & conf. type. & prob. & conf. type. & prob. \\  
\hline \hline
{\tt 000+0 } &       0.917 & {\tt 000+0 } &       0.694 & {\tt 000+0 } &       0.340 & {\tt 000-0 } &       0.694 & {\tt 000-0 } &       0.917\\
{\tt 0+0+0 } &       0.165 & {\tt 0+0+0 } &       0.102 & {\tt 000-0 } &       0.339 & {\tt 0-0-0 } &       0.102 & {\tt 0-0-0 } &       0.166\\
{\tt -0-++ } &       0.000 & {\tt 000-0 } &       0.060 & {\tt 0-0+0 } &       0.068 & {\tt 000+0 } &       0.060 & {\tt -0-++ } &       0.000\\
{\tt -0-+- } &       0.000 & {\tt 0000+ } &       0.046 & {\tt 000+- } &       0.036 & {\tt 0000- } &       0.047 & {\tt -0-+- } &       0.000\\
{\tt -0--+ } &       0.000 & {\tt 0-0+0 } &       0.039 & {\tt 000-+ } &       0.036 & {\tt 0-0+0 } &       0.039 & {\tt -0--+ } &       0.000\\
{\tt -0--- } &       0.000 & {\tt 000+- } &       0.035 & {\tt 0+0+0 } &       0.035 & {\tt 000-+ } &       0.035 & {\tt -0--- } &       0.000\\
{\tt -0-0+ } &       0.000 & {\tt 000++ } &       0.027 & {\tt 0-0-0 } &       0.035 & {\tt 000-- } &       0.027 & {\tt -0-0+ } &       0.000\\
{\tt -0-0- } &       0.000 & {\tt 000-+ } &       0.009 & {\tt 0000+ } &       0.035 & {\tt 000+- } &       0.009 & {\tt -0-0- } &       0.000\\
{\tt 00-00 } &       0.000 & {\tt 0+0++ } &       0.005 & {\tt 0000- } &       0.035 & {\tt 0000+ } &       0.006 & {\tt 00-00 } &       0.000\\
{\tt -00++ } &       0.000 & {\tt 0++00 } &       0.005 & {\tt 000++ } &       0.018 & {\tt 0-0-+ } &       0.005 & {\tt -00++ } &       0.000\\
\end{tabular}
\end{table*}

\begin{table*}
\caption{ \label{tab2}The distribution of typical configurations in the case (II) 
for $\beta J=-5$; {\em conf. type} corresponds to a configuration type (see 
Fig.\ref{fig:conf}); {\em prob.} corresponds to a probability magnitude for 
the configuration type. The symmetrical (with respect to vertical axis) 
configurations are not recognized, i.e.
{\tt 0-000} and {\tt 000-0} are the same configuration types.}
\begin{tabular}{c|l|c|l|c|l|c|l|c|l}
\multicolumn{2}{c|}{q=0} & 
\multicolumn{2}{c|}{q=0.25} & 
\multicolumn{2}{c|}{q=0.5}  & 
\multicolumn{2}{c|}{q=0.75}  & 
\multicolumn{2}{c|}{q=1.0} \\
\hline \hline
conf. type & prob. & conf. type. & prob. & conf. type. & prob. & conf. type. & prob. & conf. type. & prob. \\  
\hline \hline
{\tt 000-0 } &  0.986  & {\tt 000+0 } &  0.356 & {\tt 000+0 } &  0.340 & {\tt 000-0 } &   0.355 & {\tt 000+0 } &  0.989\\
{\tt 0000- } &  0.011  & {\tt 000-0 } &  0.312 & {\tt 000-0 } &  0.339 & {\tt 000+0 } &   0.313 & {\tt 0000+ } &  0.009\\
{\tt 000-- } &  0.003  & {\tt 0-0+0 } &  0.070 & {\tt 0-0+0 } &  0.068 & {\tt 0-0+0 } &   0.069 & {\tt 000++ } &  0.002\\
{\tt 0-0-0 } &  0.002  & {\tt 000+- } &  0.062 & {\tt 000+- } &  0.036 & {\tt 000-+ } &   0.063 & {\tt 0+0+0 } &  0.001\\
{\tt -0-++ } &  0.000  & {\tt 0-0-0 } &  0.049 & {\tt 000-+ } &  0.036 & {\tt 0+0+0 } &   0.048 & {\tt -0-++ } &  0.000\\
{\tt -0-+- } &  0.000  & {\tt 000-- } &  0.042 & {\tt 0+0+0 } &  0.035 & {\tt 000++ } &   0.043 & {\tt -0-+- } &  0.000\\
{\tt -0--+ } &  0.000  & {\tt 0000- } &  0.036 & {\tt 0-0-0 } &  0.035 & {\tt 0000+ } &   0.036 & {\tt -0--+ } &  0.000\\
{\tt -0--- } &  0.000  & {\tt 0000+ } &  0.021 & {\tt 0000+ } &  0.035 & {\tt 0000- } &   0.022 & {\tt -0--- } &  0.000\\
{\tt -0-0+ } &  0.000  & {\tt 0+0+0 } &  0.020 & {\tt 0000- } &  0.035 & {\tt 0-0-0 } &   0.020 & {\tt -0-0+ } &  0.000\\
{\tt -0-0- } &  0.000  & {\tt 000-+ } &  0.016 & {\tt 000++ } &  0.018 & {\tt 000+- } &   0.016 & {\tt -0-0- } &  0.000\\
\end{tabular}
\end{table*}

The probabilities seem to have a
very small dependence on the $\beta J$ parameter. However the $q$
parameter induces a strong influence on the configuration probability occurrence. 
No simple empirical fit has been found for such dependences. For the sake
of conciseness and argument we only display two tables (Table \ref{tab1} and \ref{tab2}) for $\beta J=-5$
and for $\beta J=5$ respectively for case II and various $q$ values.

\begin{figure*} 
\includegraphics[width=10cm]{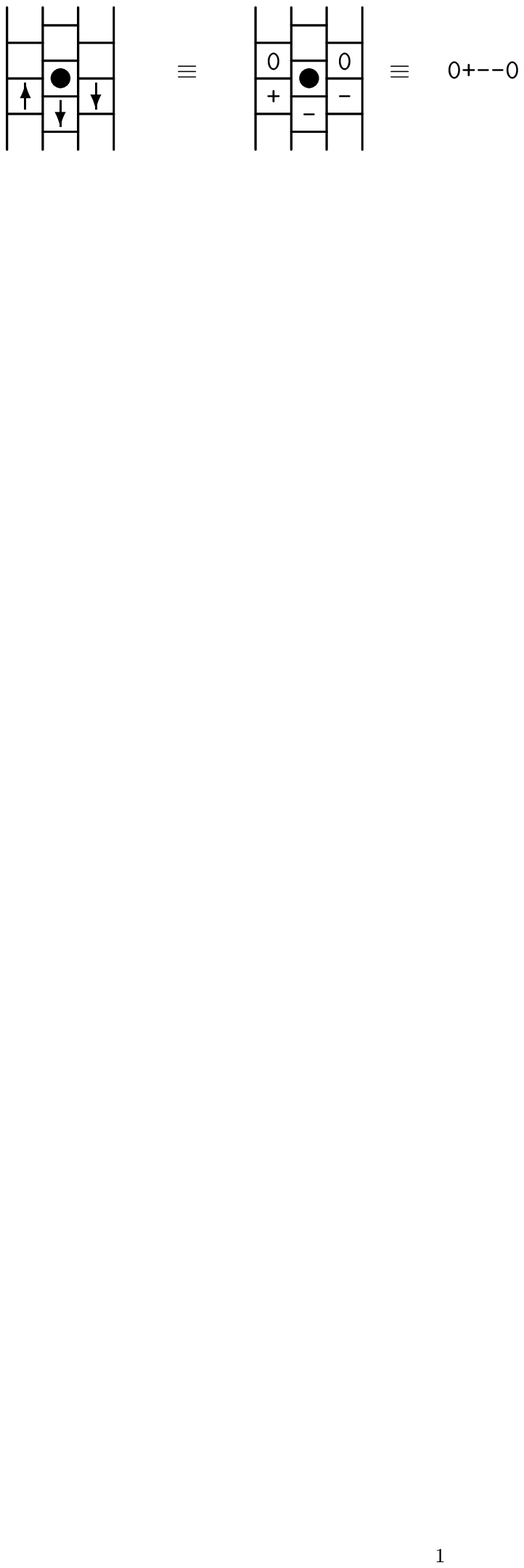} 
\caption{\label{fig:conf}The label notations for Tables \ref{tab1} and \ref{tab2}: 
the arrows denote a spin sign ($+$ is up), if on the site there is any spin; an 
empty site is $"0"$, the black circle denotes the sticking spin. On the right 
for both pictures the corresponding configuration type is written.} 
\end{figure*} 

We notice that the same configurations are most probable, in both relaxed and
non-relaxed cases. The probability of occurrence for the most frequent
configuration when $\beta J<0$ is slightly smaller for the non-relaxed case, but this
implies that the other configurations have a higher probabilities and
occur more often when there is no relaxation. This entails a larger
diversity of possible configurations in absence of relaxation
proceses. For $\beta J>0$ the probability of occurrence for 
the most frequent configuration in the non-relaxed case is not necessarily smaller 
compared to the II and III case; the differences are slender 
and may be sensitive to numerical details.

\section{ CONCLUSIONS} \label{sec:conclus}

Granular pile spreading processes are driven by cooperative non-linear
evolution rules. This leads to develop patterns which often reach a high
level of complexity. We have combined topology and mass (or size) in
order to describe granular piling in a simple way on a 1-D   substrate.
The grains are taken to be anisotropic, and the pile  forming through a
ballistic deposition process at first. However, in order to take into
considerations and observations by wonderful colleagues from the
pharmacy department\cite{delattre}, we have allowed the granular pile to somewhat
relax, in removing the  constraint of ballistic deposition when the
grain reaches the pile. We have allowed for its downward motion along
neighbourhood columns.

The pile density of course increases with relaxation.  The size distribution of clusters
follows intricate laws. Above a percolation transition, a power law
distribution is found, but it is like an exponential  law below. Such a
transition occurs at non trivial values in the ($q$, $\beta J$) plane. Slight cluster
configuration differences exist whether the ''spin-spin interaction
energy'' is positive or negative.

The above changes in density could be usefully crosschecked with new experimental studies.
This would give some information on the constact energy $\beta J$ and the spin-spin probability,
the latter through the measurment on the falling spin. Time dependent 
effects should now be usefully investigated both in simulations and in
experimentally realistic cases. For example Majumdar and Dean \cite{majumdar} found that the magnetization relaxes 
extremely slowly, in an inverse logarithmic fashion for 1-D chains. The higher dimesnionality cases should be of 
interest. The role of avalanches, as in real sand/grain piling should also be of
interest next, even taking into account the toppling of clusters of granules\cite{head}.

\begin{acknowledgments} KT is supported through an Action de Recherches
Concert\'ee Program of the University of Li$\grave e$ge (ARC 02/07-293).
\end{acknowledgments}

\end{document}